\documentclass[conference]{IEEEtran}

\usepackage{amssymb,amsmath,amsthm}
\usepackage{courier}
\usepackage{graphicx}
\graphicspath{{figures-pdf/}}
\usepackage{color}

\usepackage{cite}
\usepackage{url}

\usepackage{float}
\usepackage{multirow,bigstrut}
\usepackage{calc}   

\usepackage{algorithm}
\usepackage{algorithmic}

\usepackage{hyperref}
\hypersetup{
 linktocpage=true,
 pdfborderstyle={/S/S/W 1},
 hyperindex=true,
 bookmarks=true,
 bookmarksopen=true,
 bookmarksnumbered=true,
}

\ifCLASSINFOpdf
\else
\fi
\begin{document}
%
\title{DeepCheck: A Non-intrusive Control-flow Integrity Checking based on Deep Learning}

\author{\rm{Jiliang Zhang, Wuqiao Chen, Yuqi Niu}\\
\textit{College of Computer Science and Electronic Engineering} \\
\textit{Hunan University, China}\\
Email: zhangjiliang@hnu.edu.cn
}


%


\maketitle

\begin{abstract}
Code reuse attack (CRA) is a powerful attack that reuses existing codes to hijack the program control flow. Control flow integrity (CFI) is one of the most popular mechanisms to prevent against CRAs. However, current CFI techniques are difficult to be deployed in real applications due to suffering several issues such as modifying binaries or compiler, extending instruction set architectures (ISA) and incurring unacceptable runtime overhead. To address these issues, we propose the first deep learning-based CFI technique, named DeepCheck, where the control flow graph (CFG) is split into chains for deep neural network (DNN) training. Then the integrity features of CFG can be learned by DNN to detect abnormal control flows. DeepCheck does not interrupt the application and hence incurs zero runtime overhead. Experimental results on Adobe Flash Player, Nginx, Proftpd and Firefox show that the average detection accuracy of DeepCheck is as high as 98.9\%. In addition, 64 ROP exploits created by ROPGadget and Ropper are used to further test the effectiveness, which shows that the detection success rate reaches 100\%.
\end{abstract}


%
\IEEEpeerreviewmaketitle

\section{Introduction}

Insecure  system programming languages such as C and C++ lead to a number of vulnerabilities in the software. According to the recent security threat report \cite{ISTR}, the number of software vulnerabilities has increased significantly in the past 10 years. Attackers exploit various vulnerabilities to invade the system without authorization to steal personal privacy information, damage computer systems, spread computer viruses, and even conduct malicious blackmail.

Code Reuse Attacks (CRAs), such as return-oriented programming (ROP) \cite{Roemer2012} and jump-oriented programming (JOP) \cite{Bletsch2011}, exploit the memory overflow vulnerabilities and reuse the existing small code fragments called gadgets that end with the branch instruction such as \texttt{ret} or \texttt{jmp} to hijack the program control flow without injecting any malicious codes to perform malicious actions \cite{Roemer2012}. CRAs have been extended to different platforms \cite{Checkoway2010} such as PowerPC, Atmel AVR, SPARC, Harvard and ARM, and shown the powerful attack on well-known commercial software such as Adobe Reader \cite{SU}, Adobe Flash Player \cite{SA} and QuickTime Player \cite{AQ}.

Address Space Layout Randomization (ASLR) and Control Flow Integrity (CFI) are two mainstream defense techniques against CRAs. ASLR randomizes the code offset in virtual memory so that attackers are difficult to derive the addresses of gadgets to hijack the control flow. ASLR includes instruction location randomization \cite{Hiser2012}, code pointer randomization \cite{Lu2015}, function randomization \cite{Fu2016}, and real-time code page address randomization \cite{Chen2017}. However, the security of ASLR is affected by the randomized entropy, and the randomized addresses can be guessed by brute-force attacks \cite{Liu2011}. In addition, ASLR can be bypassed with advanced CRAs, such as JIT-ROP \cite{Snow2013} and COOP \cite{Schuster2015}. The CFI security policy dictates that software execution must follow the paths of a control flow graph (CFG) determined ahead of time \cite{Abadi2005}. Current CFI techniques can be classed into software-based \cite{Abadi2005,Budiu2006,Niu2014,Kuznetsov2014} and hardware-assisted \cite{Christoulakis2016,Davi2015,Sullivan2016,Qiu2018} CFI. Software-based CFI inserts special detection instructions into the binary file \cite{Abadi2005} or adds a runtime monitoring module \cite{Niu2014,Niu2015,Kuznetsov2014} to detect abnormal control flows without the special hardware support and hence it is easy to deploy. However, the software-based CFI need to modify the binary code or the compiler and incurs high runtime overhead. Recently, hardware-assisted CFI has attracted much attention as it can reduce the runtime overhead largely by introducing special control flow checking instructions \cite{Davi2015,Mashtizadeh2014,Cheng2014,VanderVeen2015} or additional hardware modules \cite{Davi2014a,Das2016}. However, hardware-assisted CFI suffers from the compatibility issue because of the requirement of ISA extension, compiler modification and/or special hardware support. The advantages and disadvantages of current defenses are summarized in Table~\ref{tab:1}.

\begin{table*}[!htbp]
\newcommand{\tabincell}[2]{\begin{tabular}{@{}#1@{}}#2\end{tabular}}
\centering
\caption{Advantages and disadvantages of current defense mechanisms.}
\begin{tabular}{c|c|c|c}
  \hline
  \textbf{Techniques}    &  \textbf{ASLR} & \textbf{Software-based CFI}  &   \textbf{Hardware-assisted CFI}  \\ \hline
  Advantages & Easy to deploy & Easy to deploy & Low runtime overhead \\ \hline
  Disadvantages & Easy to bypass & \tabincell{c}{High runtime overhead \\ Binary rewriting}
  & \tabincell{c}{ISA extension \\ Compiler modification \\ Special hardware support}
  \\ \hline
\end{tabular}
\label{tab:1}
\end{table*}

This paper proposes the first deep learning-based CFI technique, named DeepCheck. DeepCheck does not need to modify ISAs and compilers and will not incur extra runtime overhead, thus it has good compatibility and practicability. We have implemented the prototype of DeepCheck on Linux. The experimental results on four commercial applications show that DeepCheck has high accuracy (98.9\%), low false positive rate (0.15\%) and low false negative rate (0.60\%). The real CRA payloads are further utilized to test the effectiveness of DeepCheck, which shows that the detection success rate is up to 100\%.

The source code to reproduce our experiment is available online at \textcolor[rgb]{0.00,0.00,1.00}{http://hardwaresecurity.cn/DeepCheckCode.zip}.

The rest of this paper is organized as follows. Section II introduces some related definitions, concepts and terminologies. Section III gives a detailed introduction about the proposed DeepCheck. Experimental results and analysis are reported in Section IV. Limitations of DeepCheck are discussed in Section V. Related work is elaborated in Section VI. Finally, we conclude in Section VII.

\IEEEpubidadjcol 

\section{Background}
\label{sec:networkFixed}

In this section, we provide a brief overview of the technical concepts we use in the rest of this paper and give a detailed explanation when necessary.

\subsection{Code Reuse Attacks}

There are many gadgets in a binary executable file. Each gadget can complete a small step for a complete attack, such as changing the value in a fixed register or calling a system function. CRAs can hijack the control flow by finding the existing gadgets with different functions in the original binary and chain them to implement the achieve attack instead of injecting malicious codes. Advanced CRAs differ in the use of gadgets, e.g., COOP \cite{Schuster2015} regards a function as a gadget.

ROP and JOP are two representative CRAs. The ROP attack utilizes the gadget ending with the ret instruction. The attacker chains gadgets by overwriting the original return address of the stack with the address of the required gadget through the program vulnerability such as buffer overflow. When the \texttt{ret} instruction is executed, the CPU will get the address of the gadget instead of the original address of the program. In this way, the attacker can hijack the control flow of program. The JOP attack uses indirect \texttt{jmp} instead of \texttt{ret} instruction to change the control flow. The attacker uses a dispatcher gadget as a control unit to implement control flow transfer from one gadget to another. Since the destination address of the \texttt{jmp} instruction is stored in the register, attackers need to modify the register indirectly by modifying the contents of stack to chain gadgets.

\subsection{Control Flow Integrity}

\begin{figure}[!htb]
  \centering
  \includegraphics[width=\columnwidth]{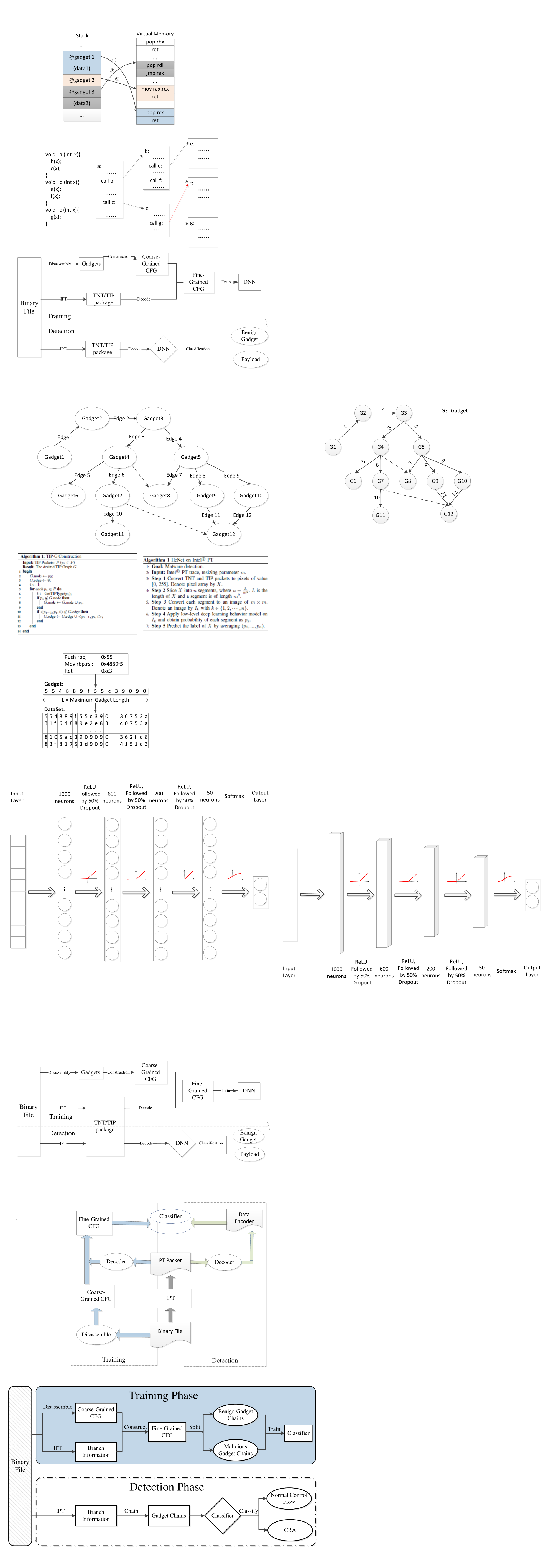}\\
  \caption{An example of a forward CFI}
  \label{fig:1}
\end{figure}

Attackers utilize memory vulnerabilities to hijack the control flow in order to execute the code in any other location in memory. CFI forces the control flow of a program to follow the CFG by verifying the execution of branch instructions. Before running the program, CFI analyzes the source program or binary code to derive the normal execution path of the program and generates a CFG. The security of CFI depends on the accuracy of the CFG. CFI with fine-grained CFG is more secure than coarse-grained CFI. However, all static analysis methods that statically analyze binary files can only build the coarse-grained CFG \cite{Burow2017}.

Control flow transfer can be classed into the forward and backward. The forward control flow is transferred by branch instructions such as \texttt{call} and \texttt{jmp}. The backward control flow is transferred by the \texttt{ret} instruction. Fig.~\ref{fig:1} shows an example of a forward CFI. An attacker attempts to change the control flow from function c to function f. However, since there is no direct edge from c pointing to f in CFG, such malicious action will be prevented by CFI.

\begin{figure*}[!htb]
  \centering
  \includegraphics[width=0.8\textwidth]{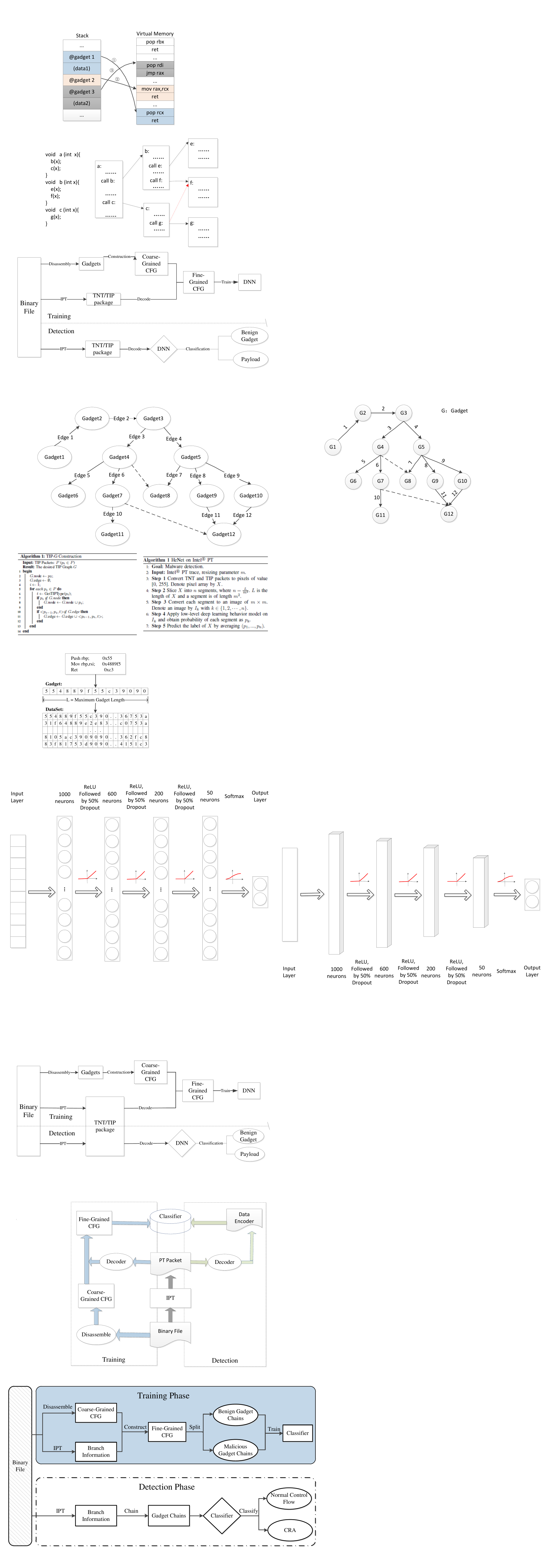}\\
  \caption{The framework of DeepCheck}
  \label{fig:2}
\end{figure*}

\subsection{Intel Processor Trace}

Intel Processor Trace (IPT) \cite{IPT} is a new hardware feature introduced in the Intel Core M 5th-generation processor for software debugging and performance analysis. IPT tracks the execution of each application dynamically and records control flow information in packets. Then these packets will be compressed and stored in the physical memory. Since the packets are not cached, the performance overhead of IPT is low. Taken Not-Taken (TNT) packets, Target IP (TIP) packets, and Flow Update Packets (FUP) are used for control flow tracking. The TNT collects the conditional branch information and records its two states (Taken (1) or Not Taken (0)) with 1-bit. TIP records the destination address of indirect branches (such as \texttt{callq}, \texttt{ret}, \texttt{jmpq}). Asynchronous events such as exceptions and interrupts are saved in FUP.

IPT can track the entire execution of one or more programs. Since the data packet only records the target address information, it is essential to combine the recorded data packet with the binary information to construct a complete CFG. The administrator can set the IPT privilege to the kernel level or user level by configuring the specified parameters, which can track the control flow in different states. DeepCheck is used to protect the user program. Hence, we set the IPT to the user level tracking.

\section{The Proposed DeepCheck}
\label{sec:networkFloat}
We assume that some basic security mechanisms have been deployed in the system, such as data execution prevention (DEP). Therefore, the attacker cannot inject malicious code. An attacker can read/write data segments arbitrarily, but can only read/execute code segments. We also assume that the application is trusted. However, an attacker can exploit program vulnerabilities, such as buffer overflows, to obtain information about any memory location. In addition, the program can't generate code dynamically or contain self-modifying code so that the CFG obtained by static analysis is correct. Most applications can meet this assumption.

\subsection{The Framework of DeepCheck}

The basic framework of our proposed DeepCheck is shown in Fig.~\ref{fig:2}. DeepCheck includes the training phase and detection phase. In the training phase, the binary file is disassembled to build a CFG where each node indicates a gadget. However, such static analysis method can only construct a coarse-grained CFG. Therefore, IPT is used to monitor the control flow to collect address information which is combined with the coarse-grained CFG to build a fine-grained CFG. Then, the CFG that consists of gadgets is split to meet the requirement of the input data for neural network training. In a CFG, two gadgets are connected by one edge. Each chain split from the CFG is a correct control flow trace. These chains constitute a benign training sample of the neural network model. The gadgets that have no edges in the CFG can be connected to build malicious samples because CRAs will connect some nodes that are not connected in the CFG to violate normal control flow.

Since neural network models can only recognize input data in a fixed format, we propose to encode the split gadget chains with the coding method introduced in section C.

In the detection phase, IPT tracks each indirect branch instruction during the program execution and records the destination address in packets. Then the packets are parsed and the destination address information is matched with the encoded gadgets to construct the test data. Finally, the data are submitted to the trained neural network for detection. If the gadget chain is classified as an illegal control flow transfer, the program will be interrupted. Specifically, DeepCheck collects IPT packets from the memory buffer in real time. It is necessary to combine the obtained address with the binary files to derive the instruction information related to the control flow transfer which requires to be encoded for DNN detection. However, it will bring high performance overhead. To address this issue, in the training phase, we calculate the offset of the target address from the binary base address, and the offset is combined with the generated gadget code to form an [offset, gadget] pair. Therefore, in the detection phase, the encoded data for the gadget can be obtained by using the address information in IPT packets. When the program performs an indirect control flow transfer, the gadget at the indirect branch is combined with the gadget at its destination address to form a gadget chain. Then the chain is input to the trained classifier for detection. If the chain is detected as a negative sample, a CRA will be detected.

The reason for constructing negative samples based on the original CFG is: 1) A large number of data samples are required for training the neural network. However, it is difficult to get sufficient training samples in the real world. In addition, it is arduous to collect different types of samples because the payload that is used to build CRAs has myriad variations. Fortunately, if CRAs are parsed from the control flow transfer, CRA variations can be controlled in a range. The original CFG contains different types of gadgets that can be utilized to conduct CRAs. Hence, a large number of negative samples can be constructed. 2) In CRAs, the control flow is changed from a gadget to another. In the corresponding CFG, this process is to connect a node to another node. Therefore, the negative samples constructed based on the original CFG have little difference with the real CRAs.

\subsection{Fine-grained CFG Construction}

The construction of CFG is critical for DeepCheck. Our goal is to build a fine-grained CFG to improve the detection accuracy of neural networks. Assuming that the binary file used for analysis has not been maliciously modified, that is, the constructed CFG is trusted.

CFG is obtained by disassembling and statically analyzing the executable binary. In \cite{Zhang2013,Veen2016,Zhang2015}, coarse-grained CFGs are constructed by binary analysis. BinCFI \cite{Zhang2015} generates a CFG by statically analyzing the \texttt{EIF} file. CCFIR \cite{Zhang2013} statically analyzes the \texttt{PE} file to find the target of all indirect branch instructions. TypeArmor \cite{Veen2016} uses use-def and liveness analysis to limit indirect call targets, and obtains indirect jump targets by the binary analysis framework. The collected offline CFG collected offline is coarse-grained. If the neural network is trained based on the coarse-grained CFG, a high false negative rate will be produced. Therefore, we use IPT to track the execution process of the program and obtain the runtime information by collecting the IPT package to improve the CFG accuracy.

In the packets generated by IPT, we use TNT and TIP to construct CFG. The TNT packet is used to record direct conditional branch information. The direct branch includes the general direct branch and the conditional direct branch. The general direct branch has a fixed target, and all such branch information can be derived when the binary file is analyzed statically. Therefore, only conditional direct branch information needs to be obtained by TNT. The TIP packet is used to record indirect branch information. Since the targets of indirect branch can be changed arbitrarily, it is a challenge to obtain all the information accurately. Therefore, the TIP package is collected as much as possible to make the indirect branch target accurate. Specifically, we need to generate a variety of different input data. For each set of input data, TNT and TIP packets are collected during the execution. The collected address information is combined with the coarse-grained CFG, and the legal control flow transfer path is added to the original CFG to construct a fine-grained CFG. The dataset is used to train the neural network model which can automatically learn the features of the control flow transfer. In DeepCheck, the CFG is split into many gadget chains. Each chain reflects a part of control flow features.

\subsection{CFG Splitting}

CFG is split into chains and each chain is encoded into data for training the neural network. If the CFG is directly used for training instead of being split into gadget chains, two issues will arise: 1) Difficult data collection. This paper builds CFGs from applications, and each application generates a unique CFG. If the entire CFG is utilized as input data, then massive CFGs are required to collect. However, a large number of dataset can be constructed by splitting the CFG, which solves the problem of difficult collecting data. 2) Poor detection effectiveness. If the model is trained with the entire CFG, the neural network learns the control flow features based on all applications. However, CRA changes the control flow of a particular application, which is difficult to be detected.

\begin{figure}[!htb]
  \centering
  \includegraphics[width=\columnwidth]{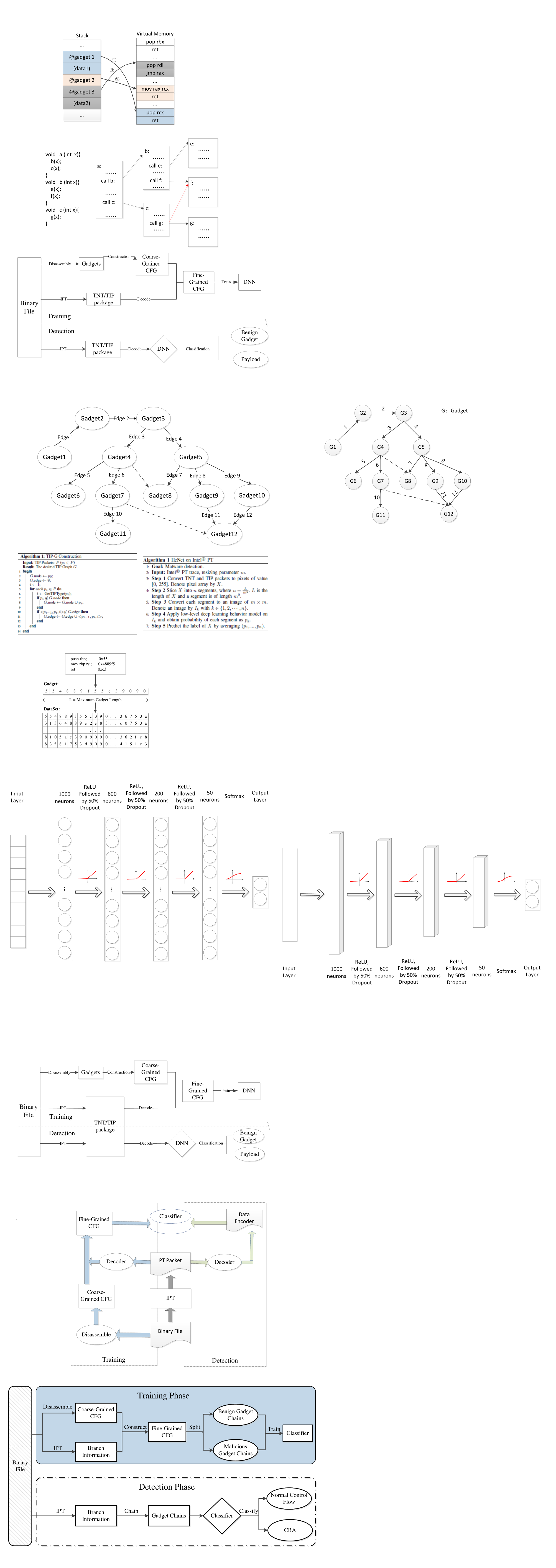}\\
  \caption{An example of partial CFG}
  \label{fig:3}
\end{figure}

As shown in Fig.~\ref{fig:3}, CFG consists of nodes and edges. Each node is a gadget. Two gadgets are connected by a directed edge, representing the path of the control flow transfer. After splitting the CFG in a single edge, Gadget1, Gadget2 and the directed edge (Edge1) form a gadget chain. Similarly, Gadget2, Gadget3 and Edge 2 can form a gadget chain. If the splitting is done in units of two edges, Gadget1, Gadget2, Gadget3, Edge 1 and Edge 2 form a gadget chain. During the execution of the program, if the control flow is transferred from Gadget4 to Gadget8 (see the dotted line), the detector considers that an attack has occurred because there is no such edge in the actual CFG.

Branch instructions are classed into direct branches and indirect branches. For indirect branches, they may have many targets and branch targets can be changed. Thus, a single edge is used as the basic unit for splitting, which ensures that the target of each indirect branch is in the benign training dataset. For direct branches, as the targets cannot be modified, two edges are used as the basic unit for splitting. The ratio of direct branches in gadget chains can be reduced by two edges (or more) to increase the proportion of indirect branches in the data set and improve the accuracy of the neural network.

Malicious gadget chains are derived by connecting gadgets that have no edge in the CFG. In this way, a large number of negative samples can be generated without manually constructing the payload. As shown in Fig.~\ref{fig:3}, Gadget3, Gadget4, and Gadget8 can form a malicious gadget chain because Gadget4 and Gadget8 are randomly connected. When the control flow is transferred from Gadget3 to Gadget4, the detection mechanism will judge that it is a normal execution flow. However, when the program performs control flow transfer from Gadget4 to Gadget8, it will be considered as an attack.

The proposed splitting algorithm is shown in Algorithm 1. The input data is CFG. The output is the gadget chains obtained by splitting the CFG. At the beginning of the algorithm, two adjacent nodes in the CFG are taken. If the first node ends with an indirect branch, the two nodes are linked into a benign gadget chain. If the last instruction of the node is a direct branch, the two nodes, together with the adjacent third node, are linked into a gadget chain. Finally, malicious gadget chains are generated by randomly connecting three unconnected nodes.

\begin{algorithm}
\renewcommand{\algorithmicrequire}{\textbf{Input:}}
\renewcommand{\algorithmicensure}{\textbf{Output:}}
\caption{Split Control Flow Graph}
\label{alg1}
\begin{algorithmic}[1]
\REQUIRE Control Flow Graph $G$
\ENSURE Benign Gadget Chains $S1$, Malicious Gadget Chains $S2$
\FOR {$g_1 \in G,g_2 \in G $}
\IF {$g_1$ is Direct branch}
\STATE get $g_3$ in $G$
\STATE $S1 \leftarrow$ $connect(g_1,g_2,g_3)$
\ELSIF {$g_1$ is Indirect branch}
\STATE $S2 \leftarrow$ $connect(g_1,g_2)$
\ENDIF
\ENDFOR
\FOR {$g_1 \in G $}
\STATE $g_1,g_2 \leftarrow$ $random(G)$
\IF {$g_1,g_2,g_3$ is not a chain in G}
\STATE $S2 \leftarrow$ $connect(g_1,g_2,g_3)$
\ENDIF
\ENDFOR
\RETURN $S1,S2$
\end{algorithmic}
\end{algorithm}

\subsection{Data Representation}

The input data of neural network must be the numerical data in a uniform format with a fixed length. If the binary instructions are used as training data directly, they need to be filled to a fixed length with 0 or 1. For example, if 0 is used to fill each instruction in the binary form of the gadget \texttt{\{push rbp; mov rbp, rsi; push rbx; ret\}} to the same length of 3 bytes, the whole data becomes a string consisting of a large number of 0 and 1. It is difficult for the neural network to distinguish the difference among the data. If one-hot encoding is adopted, all instructions in a gadget are required to be split into a single byte. For example, firstly, the gadget \texttt{\{push rbp; mov rbp, rsi; push rbx; ret\}} is split into [0x55, 0x48, 0x89, 0xf5, 0x53, 0xc3]. Then each byte is represented by a 256-bit binary string. Take 0x55 as an example, 0x55 in decimal is 85, so the 85th bit of the 256-bit binary string is 1 and the other bits are all 0, which is represented like [0, 0, 0, ..., 1, ..., 0]. However, when a lot of instructions are used to constitute a gadget, each item of data generated by the one-hot encoding method is extremely long, which increases the complexity of the input data greatly.

\begin{figure}[!htb]
  \centering
  \includegraphics[width=\columnwidth]{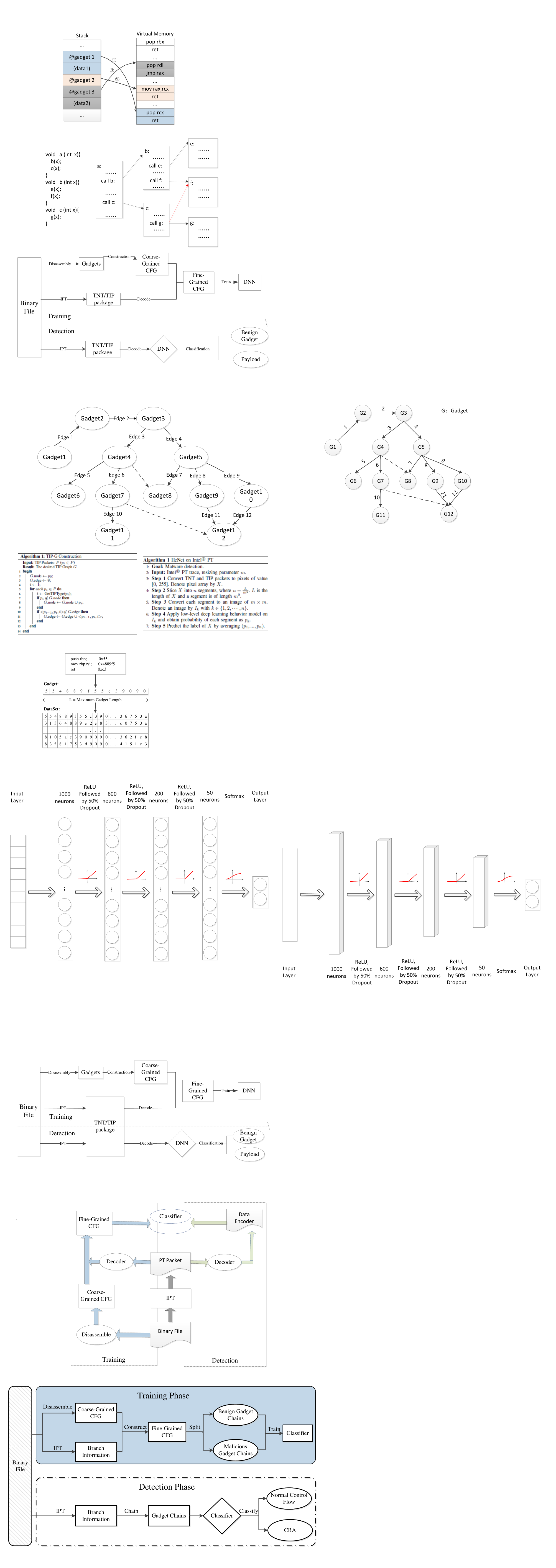}\\
  \caption{An example of data coding}
  \label{fig:4}
\end{figure}

To address the above issues, we propose to encode the gadget using the method shown in Fig.~\ref{fig:4}. Each instruction in the gadget is represented as a hexadecimal form and 4 bits are used as the basic unit for splitting. Then all the values are arranged in the order of the instructions. Since the length of the gadget is different, the \texttt{nop} command (0x90) is used to fill all of them to the same length at the end of the data, thus converting a gadget into data recognizable by the neural network. Each piece of training data can be composed of two or three gadgets. And the training data will be filled to the same length with \texttt{nop}.

\subsection{Architecture of the Deep Neural Network}

Deep Neural Network (DNN) is utilized to classify the data sets we build. Compared with traditional machine learning algorithms, DNN can get more high-dimensional features to achieve higher accuracy. Specifically, a six-layer neural network model including an input layer, an output layer, and four hidden layers is designed. The number of nodes in the four hidden layers is 1024, 512, 128, and 32, respectively. The rectified linear unit (ReLU) is used as the activation function in each hidden layer. At the same time, a 50\% dropout is set to speed up the training process and prevent overfitting. Softmax is used to convert the result to a probability value between 0 and 1.

We assume that the label of the training data is correct. Although some of the sub-chains in the execution flow of the payload are benign, it does not affect the neural network to detect the attack. In training process, the learning rate of model is set to 0.01, and the method of stochastic gradient descent is used to optimize the model. During the training of each batch, the DNN updates the weight of each layer in the network through a back-propagation algorithm and repeats the optimization process until the error converges.

\begin{table*}[!htbp]
\centering
\caption{The number of gadget chains.}
\begin{tabular}{c|c|c|c}
  \hline
  \textbf{Program} & \textbf{Vulnerability} & \textbf{Benign\ Gadget\ Chains} & \textbf{Malicious\ Gadget\ Chains}  \\ \hline
  Adobe flash 11.2.202.33 & CVE-2014-0502 & 201097 & 167580 \\
  Nginx 1.4.0 & CVE-2013-2028 & 124476 & 103980 \\
  Proftpd 1.3.0a& CVE-2006-6563 & 92042 & 60035 \\
  Firefox 3.5.10& CVE-2010-1214 & 211509 & 195920 \\ \hline
\end{tabular}
\label{tab:2}
\end{table*}

\section{Evaluation}
\label{sec:applyNetwork}
This section will evaluate the effectiveness of DeepCheck in detecting real-world CRAs from three aspects:
\begin{itemize}
  \item Can DeepCheck achieve high detection accuracy, low false positive rate, and low false negative rate when it is used to detect real applications?
  \item Is DeepCheck superior to traditional machine learning methods in terms of accuracy, false positive rate, and false negative rate?
  \item Is the runtime overhead of DeepCheck better than traditional CFI methods?
\end{itemize}

The prototype of DeepCheck has been implemented on the Ubuntu 16.04 with the kernel 4.3. \texttt{Perf} is used to configure the IPT tracking application and parse TNT and TIP packets. In addition, binary instructions are obtained by \texttt{Capstone} \cite{Capstone} and the neural network model is built by the open source framework TensorFlow. Commercial applications are used to train the neural network model. Each application trains a classifier separately. We use real payloads to evaluate the DeepCheck, and compare DeepCheck with traditional machine learning methods including SVM and logistic regression (LR). To further demonstrate the advantages of DeepCheck, the runtime overhead of DeepCheck is compared with traditional CFI mechanisms.

\subsection{Testing on Real-World Applications}

In this section, DeepCheck is evaluated with four commercial applications Adobe Flash Player, Nginx, Proftpd, and Firefox. Firstly, data is collected from these programs. To maximize the accuracy of CFG, \texttt{perf} is used to collect IPT packets and implement a decoder to extract TNT and TIP packets. Then, Algorithm1 is used to split CFG and build gadget chains. Finally, these gadget chains are encoded to generate a training data set and label both benign and malicious samples. The number of gadget chains generated by each application is shown in Table~\ref{tab:1}.

First of all, Adobe Flash Player is used to evaluate DeepCheck. 80\% of the data randomly selected from the data set is used for training, 10\% for validation, and 10\% for test. 80\% of the data set contains 80\% of benign gadget chains and 80\% of malicious gadget chains in the entire data set. Hence, the training set and validation set contain a total of 180,987 benign gadget chains and 150,822 malicious gadget chains. The test set consists of 20,109 benign gadget chains and 16,758 malicious gadget chains.

For Adobe Flash Player, the accuracy of DeepCheck is 99.2\%, the false positive rate is 0.28\%, and the false negative rate is 0.47\%. The effectiveness of DeepCheck is further evaluated on the other three real applications. For each application, 80\% of the data set is chosen as training data, 10\% for validation, and 10\% for testing. The evaluation results are shown in Table~\ref{tab:3}. The average accuracy, average false positive rate and average false negative rate of DeepCheck are respectively 98.9\%, 0.15\%, and 0.6\%. The experimental results show that the neural network trained by the real-world application can get high detection rate, low false positive rate and false negative rate. Therefore, DeepCheck learns the features of CRAs well. The reason is that DeepCheck is able to find complex features in large data sets, rather than over-fitting small training data and losing key features.

\begin{table*}[!htbp]
\centering
\caption{Results of different applications.}
\begin{tabular}{c|c|c|c|c|c}
  \hline
  \textbf{Program} & \textbf{Total\ Dataset} & \textbf{Training\ Data} & \textbf{False\ Positive} & \textbf{False\ Negative } & \textbf{Precision}\\ \hline
  Adobe flash & 368677 & 294941 & 0.28\% &0.47\% &99.2\% \\
  Nginx & 228756 & 183004 & 0.04\% &0.63\% &98.5\% \\
  Proftpd & 152077 & 121661 & 0.26\% &0.89\% &98.8\% \\
  Firefox & 407429 & 325943 & 0.02\% &0.41\% &99.4\% \\ \hline
  Average & / & / & 0.15\% &0.60\% &98.9\% \\ \hline
\end{tabular}
\label{tab:3}
\end{table*}

\subsection{Testing on Real Payloads}

In the above section, the data used to train DNN is generated by splitting the CFG. However, these data are different from real CRA exploits. In this section, we evaluate DeepCheck on real payloads and prove that DeepCheck can detect real CRA exploits.

There are 4 real exploit samples in the Exploit-DB \cite{Security2018}. To further create test samples, \texttt{ROPGadget} \cite{ROPgadget} and \texttt{Ropper} \cite{Ropper} are used to generate the exploits that execute \texttt{mprotect} or \texttt{execve}. These exploits are manually modified, for example, swapping the order of multiple gadgets without changing the attack behavior, or replacing some gadgets with new gadgets which have the same effect. Finally, we get 64 real payloads. The experimental result shows that the neural network model classifies them as real CRA payloads correctly, which further proves the feasibility of our method based on CFG splitting in detecting CRAs.

\subsection{Comparison}

Traditional machine learning methods have been proposed for ROP detection. For example, David \emph{et al}. \cite{Pfaff2015} statically detected ROP via SVM. \cite{Usui2016} detects ROP using the hidden Markov model. In this section, DeepCheck is compared with SVM and LR. Specifically, SVM with radial basis function (RBF) kernel and LR are used as classifiers, and 80\% of Firefox data set is used for training. The results are shown in Table~\ref{tab:4}.

\begin{table}[!htbp]
\centering
\caption{Comparison of DNN, SVM and LR.}
\begin{tabular}{c|c|c|c}
  \hline
  \textbf{Method} & \textbf{False\ Positive} & \textbf{False\ Negative} & \textbf{Precision}  \\ \hline
  LR & 7.1\% & 21.3\% & 82.4\% \\
  SVM & 8.7\% & 24.5\% & 74.6\% \\
  DNN & 0.15\% & 0.60\% & 98.9\% \\ \hline
\end{tabular}
\label{tab:4}
\end{table}

Experimental results indicate that DeepCheck is superior to SVM and LR in terms of the accuracy, false positive rate and false negative rate, which proves that neural networks have significant advantages in learning complex data sets and can better capture these spatial features than SVM and LR.

In addition, DeepCheck is compared with several traditional CFI methods. The evaluation metrics include security level, ISA extensions, compiler modification, runtime overhead, and non-intrusive detection. The security level includes \cite{Zhang2018}:

Level I: Only defend against ROP;

Level II: Defend against ROP and JOP;

Level III: Defend against ROP, JOP and some advanced CRAs;

Level IV: Defend against all potential CRAs.

\begin{table*}[!htbp]
\newcommand{\tabincell}[2]{\begin{tabular}{@{}#1@{}}#2\end{tabular}}
\centering
\caption{Comparison of different CFI mechanisms.}
\begin{tabular}{c|c|c|c|c|c}
  \hline
  $ $ & \tabincell{c}{Security \\ Level} & \tabincell{c}{ISA \\Extensions} & \tabincell{c}{ Compiler \\ modification}\  & \tabincell{c}{Runtime \\ Overhead}\  & \tabincell{c}{Non-intrusive}\\ \hline
  \tabincell{c}{CFI\cite{Abadi2005} \\} & I & N & N & 21\% & N \\ \hline
  \tabincell{c}{CCFI \cite{Mashtizadeh2014} \\ } & I & Y & Y & 3\%-8\% & N \\ \hline
  \tabincell{c}{CodeArmor\cite{Chen2017a} \\ } & III & N & N & 6.9\% & N \\ \hline
  \tabincell{c}{HAFIX \cite{Davi2015} \\} & I & Y & Y & 2\% & N \\ \hline
  \tabincell{c}{HCIC\cite{Zhang2018} \\ } & II & N & N & 0.95\% & N \\ \hline
  \tabincell{c}{$\mu$CFI\cite{Hu2018} \\ } & III & N & N & 10\% & N \\ \hline
  \tabincell{c}{DeepCheck} & III & N & N & 0\% & Y \\ \hline
\end{tabular}
\label{tab:5}
\end{table*}

As shown in Table~\ref{tab:5}, DeepCheck does not modify the compiler or extend ISAs, and incurs zero runtime overhead. It is worth noting that DeepCheck can achieve level III security because it can learn the features of the abnormal control flow to detect CRAs. In addition, traditional CFIs usually require inserting checking instruction or implementing the runtime detection module, which incurs additional overhead to the original program. However, DeepCheck is a non-intrusive defense mechanism that does not require instrumentation or any other form of modification. It uses the control flow information generated by IPT to implement CRAs detection. Therefore, there is no additional runtime overhead. Although DeepCheck has a certain delay in detecting CRAs, it has no influence on identifying attack behavior before the attack is completed. The reason is that the attack process of CRAs is not completed in one step (multiple gadgets need to be executed). As the control flow information generated by IPT in real time, the malicious behavior of changing control flow can be detected before completing the attack.

\section{Limitations}

This section will discuss the limitations of our proposed DeepCheck.

\begin{enumerate}
    \item The data that trains the neural network is derived from a CFG generated by a specific application, which ensures accurate detection results on the application. However, we need to train a classifier for each application separately, which is a common issue with existing CFI mechanisms.

    \item Since the neural network model is trained through a specific application, the classifier may need to be retrained when the application version is upgraded. If new functions are added to the program, the corresponding gadgets are required to be added to the data set. If some instructions are deleted, their corresponding gadgets are also removed from the data set, and the offset between gadgets needs to be adjusted. For infrequent software updates, the cost of retraining the neural network is within an acceptable range.
    \item DeepCheck has a certain false negative rate because the gadget is defined widely. The gadget consists of all the instructions from the next adjacent instruction of a branch instruction to the next branch instruction, which makes the data set contain some data of non-real gadgets. These data will enable the neural network to learn the characteristics of some non-gadgets, thereby increasing the false positive rate. However, this non-heuristic way makes DeepCheck more difficult for attackers to bypass.
    \item The accuracy of the CFG directly determines the quality of the data set, which affects the detection of the neural network model. However, generating the complete CFG for a real-world software is still an open research problem \cite{Sullivan2016}, and there is no way to implement a completely accurate CFG currently. In addition, the negative samples generated by arbitrarily connecting the CFG nodes deviate from the actual payloads, which would affect the neural network model to learn the features of CRAs. In our future work, we will explore better data construction methods to improve detection accuracy, and reduce the false positive rate and the false negative rate.
  \end{enumerate}

\section{Related Work and Motivation}

In 2005, CFI is proposed to defend against CRAs \cite{Abadi2005}. This technique inserts an ID and the ID checking code for each indirect jump instruction to enforce the execution flow of the program to follow the original CFG. Abnormal control flows will be detected due to the mismatch of inserted IDs. In CFI, the indirect branch and its target are divided into equivalence classes. All targets of an indirect branch belong to the same equivalence class, indicating that the indirect branch can target any targets in the equivalence class \cite{Niu2015}. For example, the target of branch instruction A includes B, C, and D. Then B, C, and D constitute the equivalence class of A. The implementation of CFI can be coarse-grained and fine-grained. Fine-grained CFI has a separate equivalence class for each branch target. In contrast, coarse-grained CFI usually classifies the targets of all indirect branches into an equivalence class \cite{Niu2015}. Therefore, fine-grained CFI methods will incur high runtime overhead while having higher security than coarse-grained \cite{Snow2013,Schuster2015,Carlini2015}. Security, runtime overhead and compatibility are main metrics to evaluate CFI methods which can be classed into hardware-assisted and software-based methods.

\subsection{Hardware-Assisted CFI}

In 2009, Francillon \emph{et al.} \cite{Francillon2009} proposed a stack protection mechanism that adds a microprocessor to the original computer architecture to protect the return address in the stack from the modification by attackers. It requires an additional microprocessor and the modification of ISAs, which increases hardware overhead and reduces the compatibility. In \cite{Cheng2014,VanderVeen2015}, Last Branch Record (LBR) that records the most recent control flow transfers is utilized to implement CFI, which brings lower runtime overhead than the software-based CFI. However, due to the limitation of the stack capacity, LBR can only detect a small number of executed branch targets. To overcome this limitation,\cite{Yuan2015} proposed to implement a runtime monitoring unit that only records indirect control flow transfers. Nonetheless, an interrupt is triggered whenever 16 control flow transfer information is recorded, which has a great impact on the performance. Subsequently, some hardware-assisted CFI schemes that implement CFI by extending the ISAs were proposed to check abnormal control flows \cite{Davi2015,Qiu2018,Davi2014a,Sullivan2016}.

Hardware-assisted CFI ensures that indirect jumps must follow the new CFI instructions by assigning different labels to each jump instruction. In 2016, Qiu \emph{et al}. \cite{Qiu2016} proposed to protect the control flow by using physical unclonable function (PUF) responses to linearly encrypt the return address without extending ISAs and modifying compiler. However, attackers are easy to infer the encryption key by obtaining an encrypted address. In order to address this issue, LEA-AES \cite{Qiu2018} uses the AES integrated in Intel processor to encrypt the return address. However, it still needs to extend the ISAs. Later on, HCIC \cite{Zhang2018} uses the Hamming distance matching method to avoid the key leakage without changing the compiler and ISA.

Recently, IPT is used to implement the fine-grained CFI. FlowGuard \cite{Liu2017} utilizes the information of the original IPT package and assigns different weights to control flow edges by continuous training to build a fine-grained CFG. According to the pre-generated credit-labeled, the control flow edges can be divided into the slow path and fast path to improve the detection speed. PT-CFI \cite{Gu2017} exploits IPT to implement shadow stack protection for backward CFI. GRIFFIN\cite{Ge2017} implements the fine-grained CFI at the basic block level by means of a bitwise matrix. Compared with FlowGuard and PT-CFI, GRIFFIN has a finer CFI granularity.

\subsection{Software-based CFI}

In 2010, Li \emph{et al.} \cite{Li2010} proposed a compiler-based ROP defense, which eliminates the gadgets that can be exploited by attackers via rewriting the source code in the operating system kernel. In 2013, Zhang \emph{et al.} \cite{Zhang2013} proposed a compact CFI and randomization protection method, named CCFIR, which creates a space in memory where an entry is created for the target address of each branch instruction. Branch instruction of the program can only jump to the entry in this space. However, in CCFIR, the ret instruction can be returned to any call instruction that can point to the entry of any other functions, which allows attackers to construct the malicious code fragment to hijack the control flow of program. BinCFI \cite{Anand2011} proposed to extract the CFG from binary files by combining the linear and recursive disassembly. TypeArmor \cite{Veen2016} utilizes the binary analysis and considers the high-level program semantics such as the number of function parameters and the return value to improve the precision of the CFG. However, all mechanisms based on static binary rewriting can only obtain coarse-grained CFGs. ROPdefender \cite{Cheng2014} uses dynamic binary instrumentation to implement the fine-grained CFG which incurs high runtime overhead.

In recent years, machine learning (ML), especially deep learning, has shown great advantages in autonomous vehicles, image/speech recognition, robotics, network security, and natural language processing (NLP). ML has been used to learn the feature of CRAs for CRA detection \cite{Pfaff2015,Elsabagh2017}. However, they are different from our work. In \cite{Pfaff2015}, Pfaff \emph{et al.} collect micro-architectural events generated by the hardware performance counter and use traditional Support Vector Machine (SVM) algorithm to train a classifier to realize ROP detection, which interrupts the application irregularly and incurs 5\% performance overhead. In \cite{Elsabagh2017}, Elsabagh \emph{et al.} collect microarchitecture-independent program characteristics and use unsupervised statistical model as a classifier to detect ROP. However, the detection accuracy is only 81\%. In 2018, deep learning was proposed for CRA detection \cite{Li2018}. It detects the ROP using the address space layout guided disassembly, which treats the input data as a memory address and randomly searches for the instruction sequences that may be gadgets to generate a large number of benign data sets. Then ROP detection is completed by training a convolutional neural network without a pooling layer. However, such scheme can only detect ROP attacks generated by HTTP requests, PDF files, and images, and the detection range is limited. This paper proposes the first CFI checking technique based on deep learning to prevent against all CRAs and implements such CFI strategy by splitting the fine-grained CFG. DeepCheck does not modify ISAs, compilers and binary files. Furthermore, it has high detection accuracy, good compatibility and zero runtime overhead.

\section{Conclusion}
This paper proposes the first deep learning-based CFI checking method, called DeepCheck, to resist CARs. DeepCheck has addressed several issues that traditional CFI suffered, such as extending ISAs, modifying compilers and incurring high runtime overhead. DeepCheck uses IPT to build fine-grained CFGs and then splits CFGs for neural network training. When the program is running, the branch information is acquired in real time and detected by the neural network model. The experimental results show that DeepCheck has high accuracy (98.9\%), low false positive rate (0.15\%) and false negative rate (0.60\%). In addition, it can successfully detect the real ROP exploits with a detection success rate of up to 100\%. DeepCheck is a practical non-intrusive CRA detection method with zero-runtime overhead. Therefore, compared with previous CFI methods, DeepCheck has huge advantages. In future work, we will train a more efficient neural network model to further improve its detection performance.






%
\bibliographystyle{IEEEtran_doi}
\bibliography{DeepCheck}

\end{document}